%Paper: hep-th/9406141
%From: "Costas Efthimiou" <costas@hepth.cornell.edu>
%Date: Tue, 21 Jun 94 20:09:22 EDT

%%%%%%%%%%%%%%%%%%%%%%%%%%%%%%%%%%%%%%%%%%%%%%%%%%%%%%%%%%%%%%%%%%%%%
%%%%%%%%%%%%%%%%%%%%%%%%%%%%%%%%%%%%%%%%%%%%%%%%%%%%%%%%%%%%%%%%%%%%%%%%%
%%%%
%%%%           FORM FACTORS OF 2-D INTEGRABLE MODELS
%%%%           USING RADIAL QUANTIZATION
%%%%
%%%%           talk given at the MRST meeting, held at McGill University
%%%%
%%%%                    by COSTAS EFTHIMIOU
%%%%
%%%%
%%%%%%%%%%%%%%%%%%%%%%%%%%%%%%%%%%%%%%%%%%%%%%%%%%%%%%%%%%%%%%%%%%%%%
%%%%
%%%%            Macros needed:  harvmac.tex and epsf.tex
%%%%
%%%%            Includes four figures.
%%%%            Strip from bottom of this file
%%%%            and place in  files called
%%%%                                           temporal.eps
%%%%                                           light.eps
%%%%                                           radial.eps
%%%%                                           angular.eps
%%%%            Figures start at line  1045.
%%%%
%%%%%%%%%%%%%%%%%%%%%%%%%%%%%%%%%%%%%%%%%%%%%%%%%%%%%%%%%%%%%%%%%%%%%
%%%%
\input harvmac
\input epsf.tex

\def\zdag{Z^\dagger}

\def\mass{{\rm m }}
\def\ma{\mass}
\def\ha{ {\scriptstyle{\inv{2}} }}
\def\tha{ {\scriptstyle{\frac{3}{2}} }}

\def\betah{{\hat{\beta}}}
\def\hp{\CH_\CP}
\def\hf{\CH_\CF}

\def\psib{\bar{\psi}}
\def\psip{\psi^+}
\def\psim{\psi^-}
\def\psibp{\psib^+}
\def\psibm{\psib^-}

\def\vphi{\varphi}
\def\bh{\hat{b}}
\def\bhp{\bh^+}
\def\bhm{\bh^-}
\def\sqm{\sqrt{\ma}}
\def\ez{e^{\ma z u + \ma \zb /u}}
\def\squ{\sqrt{u}}

\def\bb{\bar{b}}
\def\bbp{\bar{b}^+}
\def\bbm{\bar{b}^-}
\def\du{ \frac{du}{2 \pi i |u|} }
\def\dua{ \frac{du}{2 \pi i u} }

\def\lvacp{\langle +\ha \vert}
\def\lvacm{\langle -\ha \vert}

\def\lvacpm{\langle \pm \ha \vert}
\def\rvacpm{\vert \pm \ha \rangle}
\def\rvacmp{\vert \mp \ha \rangle}
\def\rvacp{\vert + \ha \rangle}
\def\rvacm{\vert - \ha \rangle}
\def\va#1{\vert {#1} \rangle}
\def\lva#1{\langle {#1} \vert}

\def\dstate{ {}^{\ep_1 \cdots \ep_n} \lva{\th_1 \cdots \th_n} }
\def\tstate{ {}^{+-} \lva{\th_1 , \th_2 } }

\def\psipm{\psi^\pm}
\def\psibpm{\psib^\pm }
\def\bhpm{\bh^\pm}
\def\bpm{b^\pm}
\def\bbpm{\bar{b}^\pm}

\def\bhpm{\bh^\pm}
\def\cL{\CC^{L}_{\vphi}}
\def\cR{\CC^{R}_{\vphi} }

\def\bar{\overline}
\def\hat{\widehat}
\def\*{\star}
\def\[{\left[}
\def\]{\right]}
\def\({\left(}		
\def\){\right)}

\def\zb{{\bar{z} }}
\def\frac#1#2{{#1 \over #2}}
\def\inv#1{{1 \over #1}}

\def\d{\partial}

\def\rvac{\hbox{$\vert {\rm phys} \rangle$}}
\def\lvac{\hbox{$\langle {\rm phys} \vert $}}
\def\2pi{\hbox{$2\pi i$}}

\def\dsl{\raise.15ex\hbox{/}\kern-.57em\partial}
\def\Dsl{\,\raise.15ex\hbox{/}\mkern-.13.5mu D}
\def\th{\theta}		
		\def\Ga{\Gamma}

\def\ep{\epsilon}
\def\la{\lambda}	
\def\de{\delta}		
\def\om{\omega}		
	
\def\vphi{\varphi}
		\def\CC{{\cal C}}
		\def\CF{{\cal F}}
	\def\CH{{\cal H}}	
		
		\def\CO{{\cal O}}
\def\CP{{\cal P}}

\def\rvac{\hbox{$\vert 0\rangle$}}
\def\lvac{\hbox{$\langle 0 \vert $}}

\def\2pi{\hbox{$2\pi i$}}

\def\dsl{\raise.15ex\hbox{/}\kern-.57em\partial}
\def\Dsl{\,\raise.15ex\hbox{/}\mkern-.13.5mu D}
\font\numbers=cmss12
\font\upright=cmu10 scaled\magstep1
\def\stroke{\vrule height8pt width0.4pt depth-0.1pt}
\def\topfleck{\vrule height8pt width0.5pt depth-5.9pt}
\def\botfleck{\vrule height2pt width0.5pt depth0.1pt}
\def\Zmath{\vcenter{\hbox{\numbers\rlap{\rlap{Z}\kern
0.8pt\topfleck}\kern
2.2pt
                   \rlap Z\kern 6pt\botfleck\kern 1pt}}}
\def\Qmath{\vcenter{\hbox{\upright\rlap{\rlap{Q}\kern
                   3.8pt\stroke}\phantom{Q}}}}
\def\Nmath{\vcenter{\hbox{\upright\rlap{I}\kern 1.7pt N}}}
\def\Cmath{\vcenter{\hbox{\upright\rlap{\rlap{C}\kern
                   3.8pt\stroke}\phantom{C}}}}
\def\Rmath{\vcenter{\hbox{\upright\rlap{I}\kern 1.7pt R}}}
\def\Z{\ifmmode\Zmath\else$\Zmath$\fi}
\def\Q{\ifmmode\Qmath\else$\Qmath$\fi}
\def\N{\ifmmode\Nmath\else$\Nmath$\fi}
\def\C{\ifmmode\Cmath\else$\Cmath$\fi}
\def\R{\ifmmode\Rmath\else$\Rmath$\fi}
\def\Zmath{Z}

\def\Griffin{P. Griffin, Nucl. Phys. B334, 637.}
\def\MSS{B. Schroer and T. T. Truong, Nucl. Phys. B144 (1978) 80 \semi
E. C. Marino, B. Schroer, and J. A. Swieca, Nucl. Phys. B200 (1982) 473.}

\def\VVstar{B. Davies, O. Foda, M. Jimbo, T. Miwa and A. Nakayashiki,
Commun. Math. Phys. 151 (1993) 89;
M. Jimbo, K. Miki, T. Miwa and A. Nakayashiki,
Phys. Lett. A168 (1992) 256.}

\def\form{F. A. Smirnov, {\it Form Factors in Completely Integrable
Models of Quantum Field Theory}, in {\it Advanced Series in Mathematical
Physics} 14, World Scientific, 1992.}
\def\BPZ{A. A. Belavin, A. M. Polyakov, and A. B. Zamolodchikov,
Nucl. Phys. B241 (1984) 333.}

\def\ZZ{A. B. Zamolodchikov and Al. B. Zamolodchikov, Annals
Phys. 120 (1979) 253.}
\def\Colemani{S. Coleman, Phys. Rev. D 11 (1975) 2088.}
\def\TI{H. Itoyama and H. B. Thacker, Nucl. Phys. B320 (1989) 541.}
\Title{CLNS 94/1289,
        hep-th/9406141}
{\vbox{\centerline{ FORM FACTORS OF 2-D INTEGRABLE MODELS}
       \vskip 2pt
       \centerline{USING RADIAL QUANTIZATION$^*$}
      }
}

\centerline{
             {\bf COSTAS J. EFTHIMIOU}$^\dagger$
           }
\centerline{Newman Laboratory of Nuclear Studies}
\centerline{Cornell University}
\centerline{Ithaca, NY 14853-5001}

\vskip .3in

\centerline {\bf ABSTRACT}
{
We
review some ideas from  a recent construction which
introduced the notion of vertex
operators and form factors as vacuum expectation values of
related vertex operators in the space of fields.  The vertex operators
are constructed explicitly in radial quantization.
These ideas are explained at the free-fermion point of the
sine-Gordon theory.
}

\vskip .6in

\hrule
\vskip .1in
$^*$Tallk presented  at the MRST meeting, held at McGill
University.
\hfill\break\indent
$\dagger$ ~e-mail address: costas@hepth.cornell.edu

\Date{June, 1994}

\noblackbox

\def\ot{\otimes}
\def\zb{{\bar{z}}}

\def\zbar{{\bar{z}}}

\newsec{Introduction}

For relativistic quantum field theories with a massive particle
spectrum, the main dynamical properties one is interested in are
the S-matrix, the form factors of all local fields, and the
Green's functions of these fields.   For the integrable quantum
field theories in two space-time dimensions, some of these properties
have been computed exactly.  The algebraic structures that
characterize the S-matrices are well-known \ref\rzz{\ZZ}, and
have been used to compute the S-matrices for a wide variety of models.
Bootstrap axioms satisfied by the form factors have been
formulated \ref\rkw{M. Karowski and P. Weisz, Nucl. Phys. B139 (1978)
445.}\ref\rform{\form}.  Important progress in solving the
bootstrap for the multiparticle form factors was made by
Smirnov \ref\rffold{F. A. Smirnov, J. Phys. A: Math. Gen. 19 (1986)
L575.}\rform, where he computed exactly the form factors of
certain basic fields,  such as the energy-momentum tensor and
global conserved currents, in the sine-Gordon (SG) model,
$SU(N)$ Thirring model, and $O(3)$ non-linear sigma model.
(For a recent review on the bootstrap program the reader may consult
\ref\GM{G. Mussardo, Phys. Rep. 218 (1992) 215.}.)

It was of interest to develop a more algebraic framework for the
computation of form factors, with the aim of constructing
the solutions for the complete set of fields in an
efficient manner.  A deeper understanding of such algebraic
structures is likely to facilitate generalizations to other
models, and could lead to some much-needed new
approaches to the problem of computing Green's functions.
In the works
\ref\lec{A. LeClair, {\it Spectrum Generating Affine Lie Algebras
in Massive Field Theory}, hep-th/9305110, to appear in Nucl. Phys. B.}
\ref\eflec{C. Efthimiou and A. LeClair, {\it Particle Field Duality and
Form Factors from Vertex operators}, CLNS 93/1263, hep-th 9312121}
\ref\rluk{S. Lukyanov, {\it Free Field Representation for Massive
Integrable Models},
{\it Correlators of the Jost
Functions in the Sine-Gordon Model}, Rutgers preprints
RU-93-30, RU-93-55.}
two new approaches to the computation of form factors were proposed
and were further elaborated.
In \lec\eflec, structures in radial quantization were used to construct
form factors explicitly as vacuum expectations of vertex operators
in momentum space.
The presentation of the central ideas underlying these constructions
is the subject of this talk. Hopefully by the end of the talk
it will be clear what the benefits of the new approaches are and where the
subtle points  lie.

\newsec{Conventional Multi-Particle Fock Space}

In a quantum field theory with a spectrum of massive particles,
one deals with the space of multiparticle states $\hp$.
In the context of the integrable theories in two dimensions,
we can describe $\hp$ as follows.  Introduce the so-called
Faddeev-Zamolodchikov operators $\zdag_\ep (\th )$ and
$Z^\ep (\th )$, where $\th$ is the rapidity parametrizing
the energy-momentum ($E= \ma \cosh \th ,~ P = \ma \sinh \th $),
and $\ep$ is an isotopic index, satisfying:
\eqn\eIIi{\eqalign{
Z^{\ep_1} (\th_1 ) \> Z^{\ep_2} (\th_2 )
&= S^{\ep_1 \ep_2}_{\ep'_1 \ep'_2} (\th_{12} ) ~
Z^{\ep'_2} (\th_2 ) \> Z^{\ep'_1} (\th_1 )
\cr
\zdag_{\ep_1} (\th_1 ) \> \zdag_{\ep_2} (\th_2 )
&= S^{\ep'_1 \ep'_2}_{\ep_1 \ep_2} (\th_{12} ) ~
\zdag_{\ep'_2} (\th_2 ) \> \zdag_{\ep'_1} (\th_1 )
\cr
Z^{\ep_1} (\th_1 ) \zdag_{\ep_2} (\th_2 ) ~&= ~
S^{\ep'_2 \ep_1}_{\ep_2 \ep'_1} (\th_{21} ) ~
\zdag_{\ep'_2} (\th_2) Z^{\ep'_1} (\th_1 )
+ \delta^{\ep_1}_{\ep_2} ~  \delta (\th_1 - \th_2 ) . \cr }}
Above, $S$ is the S-matrix, and $\th_{12} = \th_1 - \th_2$.
 Define particle states as follows:
\eqn\eIIii{\eqalign{
\va{\th_1 \cdots \th_n }_{\ep_1 \cdots \ep_n }
&= \zdag_{\ep_1} (\th_1 ) \cdots \zdag_{\ep_n} (\th_n ) \rvac
\cr
{}^{\ep_1 \cdots \ep_n } \lva{\th_1 \cdots \th_n}
&= \lvac Z^{\ep_1} (\th_1) \cdots Z^{\ep_n} (\th_n ) , \cr }}
where
$\rvac$ is the physical vacuum.
The space $\hp$ and its dual $\hp^*$ are spanned by the above
states:
$$\hp = \{ \oplus_n ~
\va{\th_1 \cdots \th_n }_{\ep_1 \cdots \ep_n }
\}$$
$$\hp^* = \{ \oplus_n ~
{}^{\ep_1 \cdots \ep_n } \lva{\th_1 \cdots \th_n}
\}.$$
In this space one has the completeness relation
\eqn\eIIiv{
1 = \sum_{\vec{\th}} \va{\overrightarrow{\th}} \lva{\overleftarrow{\th}}
=
\sum_{n=0}^\infty \inv{n!} \sum_{ \{ \ep_i \} } \int d\th_1 \cdots
d\th_n ~ |\th_1 ,\cdots , \th_n \rangle_{\ep_1 \cdots \ep_n}
{}^{\ep_n \cdots \ep_1 } \langle \th_n , \cdots , \th_1 | . }

\newsec{Correlation Functions and Form Factors}

In a field theory the interesting quantities are the n-point Green's
functions of the local operators
$\CO(x)$ that exist in the theory. For example, one would like to know
$$
 G(x,y)=-i\,\langle T(\CO(x)\CO(y)) \rangle ~.
$$
Using the completeness relation \eIIiv\  we may write
$${\eqalign{
G(x,y)=-i
\sum_{n=0}^\infty \inv{n!} \sum_{ \{ \ep_i \} } \int d\th_1 \cdots
d\th_n ~
 \lvac\CO(x)|\th_1 ,\cdots , \th_n \rangle_{\ep_1 \cdots \ep_n}
{}^{\ep_n \cdots \ep_1 } \langle \th_n , \cdots , \th_1 |\CO(y)\rvac
\, \th(x^0-y^0)
&\cr
    ~+~(x\leftrightarrow y)~.~~~~~~~~~~~~~~~
&\cr
          }}$$
Obviously calculation of the above correlation function inevitably
requires calculation of the quantities
$$
 {}^{\ep_n \cdots \ep_1 } \langle \th_n , \cdots , \th_1 |\CO(y)\rvac~.
$$
The spatial dependence is trivial due to displacement invariance:
$$
 {}^{\ep_n \cdots \ep_1 } \langle \th_n , \cdots , \th_1 |\CO(y)\rvac
 = {\rm exp}\left({i\sum_{j=1}^n\, p^\mu(\th_j) y_\mu}\right)  ~
  {}^{\ep_n \cdots \ep_1 } \langle \th_n , \cdots , \th_1 |\CO(0)\rvac ~.
$$
The quantities
\eqn\formf{
 F_\CO^{\ep_n \cdots \ep_1 }(\th_n , \cdots , \th_1)\equiv
  {}^{\ep_n \cdots \ep_1 } \langle \th_n , \cdots , \th_1 |\CO(0)\rvac
          }
are known as form factors of the local operator $\CO$.
Using the form factors one can give  spectral representations of the
Green's functions \ref\BjDr{ J. D. Bjorken and S. D. Drell,
{\it Relativistic Quantum Fields}, McGraw Publishing Company, 1965.}.
In particular,
$$
  G(x,y)
  = \int_0^{+\infty}\, d\sigma^2\,
 {\rho_\CO(\sigma^2)\, \Delta(x-y;\sigma)}~,
$$
where the spectral amplitude is given by
$$
   \rho_\CO(q^2)\,\theta(q^0)=2\pi\,
\sum_{n=0}^\infty \inv{n!} \sum_{ \{ \ep_i \} }
\int d\th_1 \cdots d\th_n\,
\delta^{(2)}(q-\sum_{j=1}^np(\th_j))\,
|F_\CO^{\ep_n...\ep_1}(\th_n...\th_1)|^2~,
$$
and
$$
   \Delta(x-y;\sigma)=\int\,{d^2q\over (2\pi)^2}\,{e^{-iq\cdot (x-y)}
        \over q^2-\sigma^2+i\varepsilon}~
$$
is the free 2-point Green's function.
We must say, though, that despite the progress in the calculation of form
factors, both using our methods or various other methods \rform%
\GM\lec\eflec,
the  calculation of the
spectral representation of the correlation functions in a closed form
still remains   an open
problem.

\newsec{The Space of Fields}

Usually the form factors \formf\  are calculated in the multi-particle
Fock space $\CH_\CP$ as follows.
First, one expands the operator $\CO(0)$ in terms
of annihilation and creation operators acting on states in $\CH_\CP$;
then the action
of $\CO(0)$ on the physical vacuum $\rvac$ is easily found resulting
in a linear combination of states in $\CH_p$. Finally,
the matrix element \formf\ is obtained as a linear combination of
(trivial) matrix elements  of states in $\CH_\CP$.

Now let us consider the space of fields $\hf$.  Let $\CF$ denote the
complete space of fields, and define $\va{\Phi_i} = \Phi_i (0) \rvac$,
$\Phi_i (x)  \in \CF$.
The space $\hf$ is defined as follows
\eqn\eIIv{
\hf = \{ \oplus_{\Phi_i \in \CF}  ~~ \va{\Phi_i}  \} .}
Form factors are matrix elements of fields in the space of states
$\hp$.  The basic form factors
$\dstate\Phi \rangle$, from which the more general matrix
elements may be obtained by  crossing symmetry, are inner products
of states in $\hf$ with states in $\hp^*$.  The completeness relation
\eIIiv\ allows us to map states in $\hf$ to states in $\hp$, i.e. to
view $\va{\Phi} \in \hp$:
\eqn\eIIvi{
\va{\Phi_i} =
 \sum_{\vec{\th}} \va{\overrightarrow{\th}}
 \lva{\overleftarrow{\th}} \Phi_i \rangle .}
The intuitive simplicity of the space $\hp$ is responsible for this
conventional way of thinking about form factors.

We give now a dual description of the same form factors.  Let us
suppose that one can define a dual to the space of fields $\hf^*$
with inner product and completeness relation:
\eqn\eIIvii{\eqalign{
\lva{\Phi^i} \Phi_j \rangle &= \delta^i_j  \cr
1 &= \sum_i \va{\Phi_i} \lva{\Phi^i }  . \cr }}
Then one can map a state $\va{\vec{\th}} \in \hp$ into
$\hf$.  The dual statement is
\eqn\eIIviii{
\dstate = \sum_{\Phi_i \in \CF}
\dstate \Phi_i \rangle \lva{\Phi^i } ~~~~\in \hf^* . }

In order to use these ideas to compute form factors, we need to
introduce the notion of vertex operators.  The formula
\eIIviii\ implies that one can map states in $\hp^*$ to states
in $\hf^*$.  We call this map the `particle-field map'.
We construct this map explicitly by defining vertex
operators  $V^\ep (\th )$ as follows:
\eqn\eIIix{
\dstate = \lva{\Omega} ~ V^{\ep_1} (\th_1 ) \cdots
V^{\ep_n} (\th_n ) ~~ \in \hf^* }
where $\lva{\Omega}$ is a fixed `vacuum' state, which
will be characterized completely in a later section. The
vertex operators are distinguished from the Faddeev-Zamolodchikov
operators $Z(\th )$ since they act on completely different spaces.
However the basic algebraic relations satisfied by the $Z$ operators
continue to be satisfied by the $V$ operators.
The vertex operators $V^{\ep} (\th )$ operate in the space $\hf$:
\eqn\eIIx{
V^{\ep} (\th ) : ~~~~\hf \to \hf . }
In the sequel we will describe how to construct these vertex
operators explicitly using radial quantization.  Once the
vertex operators are constructed, the form factors
$\dstate \Phi_i \rangle$ are computed directly in the
discrete space $\hf$ using \eIIix.

\newsec{Quantization Schemes}

In order to work the above simple remarks into an efficient means
of computing form factors, one must work explicitly with the space
$\hf$.
In order to make clear our choice of radial quantization we
digress on the quantization schemes; in particular we review
the four most important quantization schemes.

$\bullet$ \underbar{Temporal quantization}: In this scheme the
quantization time-like surfaces are given by $t=const$.
\midinsert
\epsfxsize = 2in
\bigskip
\vbox{\vskip -.1in\hbox{\centerline{\epsfbox{temporal.eps}}}
\vskip .1in
{\leftskip .5in \rightskip .5in \noindent \ninerm \baselineskip=10pt
Figure 1.
Temporal quantization.
\smallskip}} \bigskip
\endinsert
The momentum operator is a kinematical operator, i.e. it moves along
the surfaces while the Hamiltonian is a dynamical operator, i.e.
it moves from one surface to another.

$\bullet$ \underbar{Light-cone quantization}: In this scheme the
quantization time-like surfaces are given by $x+t=const$. The
$P^+$ is a kinematical operator while the $P^-$ is a dynamical one.
\midinsert
\epsfxsize = 2in
\bigskip
\vbox{\vskip -.1in\hbox{\centerline{\epsffile{light.eps}}}
\vskip .1in
{\leftskip .5in \rightskip .5in \noindent \ninerm \baselineskip=10pt
Figure 2.
Light-cone quantization.
\smallskip}} \bigskip
\endinsert

$\bullet$ \underbar{Radial quantization}: In this scheme the
quantization time-like surfaces are given by the circles  $r=const$. The
Lorentz operator $L$ is a kinematical operator while the
dilation operator $D$ is a dynamical one. This is evident from the fact
that the circle $r=R_1$ coincides with the circle $r=R_2$ by
simple rescaling. Note that $D$ is not a conserved operator in general.

Now, we remark that in conformal
field theory\ref\rbpz{\BPZ}\ one deals precisely with the space
of fields, and it was argued in \lec\ that for many massive quantum
field theories, the space $\hf$ is identical to its description
in the ultraviolet conformal field theory.
It thus turns out that the space $\hf$ diagonalizes the Lorentz operator $L$,
thus it can be realized as the space of radial quantization
\ref\rfhj{S. Fubini, A. J. Hanson, and
R. Jackiw, Phys. Rev. D7 (1973) 1732.}\lec.  In radial quantization,
the space $\hf$ is associated with $r=0$, whereas the
dual space $\hf^*$ is associated with $r= \infty$.
\midinsert
\epsfxsize = 2in
\bigskip
\vbox{\vskip -.1in\hbox{\centerline{\epsffile{radial.eps}}}
\vskip .1in
{\leftskip .5in \rightskip .5in \noindent \ninerm \baselineskip=10pt
Figure 3.
Radial quantization.
\smallskip}} \bigskip
\endinsert

$\bullet$ \underbar{Angular quantization}: In this scheme the
quantization time-like surfaces are given by the semi-infinite
lines $\vphi=const$.
The dilation operator $D$ is now  the  kinematical operator
while the Lorentz operator $L$ is the dynamical operator.
\midinsert
\epsfxsize = 2in
\bigskip
\vbox{\vskip -.1in\hbox{\centerline{\epsffile{angular.eps}}}
\vskip .1in
{\leftskip .5in \rightskip .5in \noindent \ninerm \baselineskip=10pt
Figure 4.
Angular quantization.
\smallskip}} \bigskip
\endinsert

The angular quantization is the basic
incredient in Lukyanov's construction\foot{We are
ignoring a technical subtlety at this point. Angular quantization
refers to the Euclidean plane while Lukyanov is using a Minkowski
space. His quantization scheme should be called ``Rindler quantization"
as he is using Rindler coordinates which can be seen as polar coordinates
in the Euclidean formulation.
The original motivation behind his construction came
from the work \ref\vvstar{\VVstar}, where the necessary
properties of these traces were understood in the
context of lattice models.}
\rluk .
In angular quantization, since the Lorentz  operator $L$
generates shifts in $\vphi$, it plays the role of the ``Hamiltonian''.  Then,
functional integrals can be represented as traces:
\eqn\func{
\lvac ~ \CO ~ \rvac  = \frac{\int ~ D\Phi  ~ e^{-S} ~ \CO}
{\int D\Phi e^{-S} }
{}~ = ~  \frac{Tr ~ ( e^{2\pi i L} \> \CO )}{Tr ~ ( e^{2\pi i L} )} . }
The $2\pi i$ constant in the factor $\exp (2\pi i L )$
is fixed by the
$2\pi$ length of the `time' $\vphi$.  In \vvstar\rluk,
the latter constant was fixed by imposing  the right
symmetry properties of the form factors expressed as these
traces.

\newsec{Radial Realization of $\hf$}

We will consider the sine-Gordon theory, defined by the action
\eqn\eIi{
S = \inv{4\pi} \int d^2 z \(  \d_z \phi \d_\zb \phi
                 + 4\la  \cos ( \betah \phi ) \)~. }
It is well known that this model is equivalent to the massive Thirring
model
\ref\rcol{\Colemani}.
In particular, at $\hat{\beta} = 1$ it becomes  free {\it massive}
Dirac model;
the free Dirac fermion fields $\psi^\pm $,
$\psib^\pm$ carry $U(1)$ charge $\pm 1$, and their dynamics
is governed by the action
\eqn\eIIi{
S = - \inv{4\pi} \int dx dt \(
\psibm \d_z \psibp + \psim \d_\zb \psip
+ i \ma ( \psim \psibp - \psibm \psip ) \) . }

In temporal quantization, the expansion of
the fermion fields in terms of momentum space operators is
as follows:

\eqn\eIIxvib{
\left\lbrack\matrix{{\psibp(x,t)}\cr{\psip(x,t)}}\right\rbrack
 =  \sqm \int_{-\infty}^\infty
d\th ~
%e^{\th/2}
\left\{
 Z^+ (\th ) \, e^{-ip(\th ) \cdot x } \,
\left\lbrack\matrix{e^{\th/2}\cr -ie^{-\th/2}}\right\rbrack
+ \zdag_- (\th )
    \, e^{i p (\th)  \cdot x } \,
\left\lbrack\matrix{-e^{\th/2}\cr -ie^{-\th/2}}\right\rbrack
\right\}
            }

Consider now radial quantization of the free-fermion
theory\rfhj\ref\rit{\TI}\ref\rgrif{\Griffin}.  Define
the usual polar coordinates
$z = (t+ix)/2 =r/2~ \exp(i\vphi) , ~
\zbar = (t-ix)/2 = r/2~ \exp (-i\vphi )$.
One can define
two distinct sectors, the periodic (p) and
anti-periodic (a), with expansions
\eqn\eIIIxvii{
\Psi_{(a,p)}^\pm =
\left\lbrack\matrix{{\psibpm}\cr{\psipm}}\right\rbrack
= \sum_\om
 \left[ \bpm_\om ~ \Psi^{(a,p)}_{-\om - 1/2} ~+~
\bbpm_\om ~ \bar{\Psi}^{(a,p)}_{-\om -1/2}  \right] , }
where for the periodic sector $\om \in \Zmath + 1/2$, and
for the anti-periodic sector $\om \in \Zmath$.
The basis spinors can be found as solutions to the
equations of motion in radial coordinates
\rgrif\eflec ; the explicit expressions
 are not going to be of importance here.

In the quantum theory the above modes satisfy simple anti-commutation
relations:
\eqn\eIIIxxiv{
\{ b^+_\om , b^-_{\om'} \} =
\{ \bbp_\om , \bbm_{\om'} \} = \de_{\om , -\om'}
, ~~~~~\{ b_\om , \bb_{\om '} \} = 0. }
These modes diagonalize the Lorentz boost operator:
\eqn\eIIIxxiiib{
\[ L , \bpm_\om \] = -\om ~ \bpm_\om , ~~~~~
\[ L , \bbpm_\om \] = \om ~ \bbpm_\om . }
The space of radial quantization consists of Fock modules built
from these oscillators.

  In the periodic sector, the vacuum is
defined to satisfy
\eqn\eIIIxxv{
\bpm_\om \> \rvac = \bbpm_\om \> \rvac = 0 , ~~~~~\om \geq 1/2 . }
The vacuum $\rvac$ is the physical one.  One constructs
Fock modules for the periodic sector as follows:
\eqn\eIIIxxvi{
\CH^L_p = \left\{ b^{\ep_1}_{-\om_1}
b^{\ep_2}_{-\om_2} \cdots
 \rvac \right\} , ~~~~~~~
\CH^R_p = \left\{ \bb^{\ep_1}_{-\om_1}
\bb^{\ep_2}_{-\om_2} \cdots
 \rvac \right\} , }
where $\om_i \in \Zmath + 1/2 , \om_i  \geq 1/2$.

In the anti-periodic sector, due to the existence of the
zero modes, the `vacuum' states are doubly degenerate.
These vacua $\va{\pm \ha}_L$ and $\va{\pm \ha}_R$ are
characterized as follows:
\eqn\eIIIxxxiv{\eqalign{
\bpm_0 \> \va{\mp \ha}_L = \rvacpm_L , ~~~~ \bpm_0 \> \rvacpm_L = 0 ,
{}~~~~ \bpm_n \> \rvacpm_L = 0,~~~~&n\geq 1 \cr
\bbpm_0 \> \va{\mp \ha}_R = \rvacpm_R , ~~~~ \bbpm_0 \> \rvacpm_R = 0 ,
{}~~~~ \bbpm_n \> \rvacpm_R = 0,~~~~&n\geq 1 . \cr}}
The dual vacua $\lvacpm$ are defined by the inner products
\eqn\eIIIxxxv{
{}_L \lva{\mp \ha} \pm \ha \rangle_L = {}_R \langle \mp \ha \rvacpm_R = 1.}
The anti-periodic
Fock spaces are defined as
\eqn\eIIIxxxvb{
\CH^L_{a_{\pm}}
= \left\{ b^{\ep_1}_{-n_1} b^{\ep_2}_{-n_2} \cdots
\cdots \rvacpm_L \right\} , ~~~~~
\CH^R_{a_{\pm}}
= \left\{ \bb^{\ep_1}_{-n_1} \bb^{\ep_2}_{-n_2} \cdots
\cdots \rvacpm_R \right\} , }
for $n_i \in \Zmath  \geq 1$.

The space of radial quantization corresponds precisely to the
space of fields $\hf$ described in a previous section.
Namely, in the periodic sector $\hf^{(p)} = \CH^L_p \otimes
\CH^R_p$, and in the
anti-periodic sector $\hf^{(a)} = \CH^L_a \ot \CH^R_a$,
where
$\CH^{L,R}_a = \CH^{L,R}_{a_+} \oplus \CH^{L,R}_{a_-} $.
What is remarkable about this result is that the structure
of the space of fields is identical to that in the massless
conformal limit.
Arguments explaining this phenomenon were given in \lec.

We present some simple examples that we will use later.
The $U(1)$ current $J_\mu$ has components
$J_z = \psi^+ \psi^-$, $J_\zbar = \psib^+ \psib^- $.
Also of interest is the energy-momentum tensor\foot{We
normalize conserved currents such that
$Q = \inv{4\pi} \int dx \( J_z + J_\zbar \) $ is the properly
normalized conserved charge.}:
\eqn\eIIIxiii{
\eqalign{
T_{zz} &= \inv{2} \( \psi^- \d_z \psi^+  - \d_z \psi^- \psi ^+ \)
, ~~~~~
T_{\zbar\zbar} = \inv{2} \( \psib^- \d_\zbar \psib^+
- \d_\zbar \psib^- \psib^+ \)  \cr
T_{z\zbar} &= \inv{2} \( \psi^- \d_\zbar \psi^+  - \d_\zbar \psi^- \psi^+ \)
, ~~~~~
T_{\zbar z} = \inv{2} \( \psib^- \d_z \psib^+ - \d_z \psib^- \psib^+ \)
. \cr }}
Using the expansions \eIIIxvii, the asymptotic expansions of the
basis spinors as $r\to 0$ and the properties
of the vacuum \eIIIxxv, one finds
\eqn\eIIIxiv{\eqalign{
\d_z^n \psi^\pm (0) \rvac &= n! \> b^\pm_{-n-\ha} \rvac ,
{}~~~~~~\d_\zbar^n \psib^\pm (0) \rvac = n! \bb^\pm_{-n-\ha} \rvac \cr
J_z (0) \rvac &= b^+_{-\ha} b^-_{-\ha} \rvac , ~~~~~~
J_\zbar (0) \rvac = \bb^+_{-\ha} \bb^-_{-\ha} \rvac \cr
T_{zz} (0) \rvac &= \inv{2} \( b^-_{-\ha} b^+_{-\tha}
- b^-_{-\tha} b^+_{-\ha} \) \rvac , ~~~~~
T_{\zbar\zbar} (0) \rvac
= \inv{2} \( \bb^-_{-\ha} \bb^+_{-\tha}
- \bb^-_{-\tha} \bb^+_{-\ha} \) \rvac \cr
T_{z \zbar} (0)\rvac &= T_{\zbar z} (0) \rvac
= - \frac{i\ma}{2} \( b^-_{-\ha} \bb^+_{-\ha}  - \bb^-_{-\ha} b^+_{-\ha}
\) \rvac . \cr}}

The anti-periodic sector is more interesting, since here
one can access fields in the sine-Gordon theory which are
not simply expressed in terms of the fermion fields.  By studying
the operator product expansion, the following identification
was made in \lec:
\eqn\eIIIxvi{
e^{\pm i \phi (0)/2 } \> \rvac ~=~
(c\ma )^{1/4} ~ \( \rvacpm_L \ot \rvacmp_R \)
\equiv (c\ma )^{1/4} ~ \va{\pm \ha}
 , }
where $\phi (z, \zb )$ is the local SG field, and
$c$ is an undetermined dimensionless constant.
The mass dimension of $1/4$ on the RHS is fixed by
the known scaling dimension of the fields $\exp (\pm i \phi /2 )$,
which is the same as in the conformal limit.
(The constant $c$ can
be fixed by specifying the 1-point functions of these fields;
see below.)   The fields $\exp( \pm i \phi /2 )$ are non-trivial
terms of the fermions, since the bosonization
relation is $\cos\phi= ( \psi^- \psib^+ - \psib^- \psi^+ )$.
All other states in $\hf^{(a)}$ correspond to regularized products of fermion
fields and their derivatives with the basic fields
$\exp (\pm i \phi /2 )$.

\newsec{Particle-Field Maps}

The basic property that allows us to construct explicitly the
particle-field maps of the form \eIIix\ is the fact that the
space of radial quantization can be obtained from the usual
space of temporal quantization by appropriate analytic continuation
in momentum space.  Since the situation in the periodic versus
the anti-periodic sector is quite different, we  present the results
separately.

Let us combine the creation and annihilation operators of temporal
quantization into a single operator as follows.  Define the
momentum space variable $u$ as
$u = e^\th $
and define operators $\bh^\pm (u) $ as
\eqn\eIVii{\eqalign{
\bhp (u) &= 2\pi \> \zdag_- (\th ) ,
{}~~~~~~~~~~~~~~~\bhm (u) = 2 \pi \> \zdag_+ (\th )  ~~~~~~~~~~~{\rm for}
{}~~~~~~ u>0,
\cr
\bhp (u) &= 2\pi i \> Z^+ (\th - i\pi )    ,
{}~~~~~~~~\bhm (u) = 2\pi i \> Z^- (\th - i\pi )    ~~~~~~~~~{\rm for} ~ u<0
.\cr }}
Then the temporal quantization expansion \eIIxvib\ may be written as
\eqn\eIIxx{
\Psi^\pm =
\left\lbrack \matrix{{\psib^\pm}\cr{\psi^\pm}}\right\rbrack = \pm \sqm
\int_{-\infty}^\infty \du \> \bh^\pm (u) ~
\left\lbrack \matrix{{1/ \squ}\cr {-i\squ} } \right\rbrack
 ~\ez~,}

\medskip
\noindent
7.1 {\it Periodic Sector}

Let us now define the following prescription for analytic continuation
of the $u$-integral in \eIIxx:
\eqn\epercont{\eqalign{
\int_{0}^\infty \du ~ \bhpm (u) &\rightarrow
 \oint_{\CC_<} \dua \( \bpm_< (u) + \bbpm_< (u) \)  \cr
\int_{-\infty}^0 \du ~ \bhpm (u) &\rightarrow
 \int_{\CC_> } \dua \( \bpm_> (u) + \bbpm_> (u) \)  , \cr}}
where
the contour $\CC_<$ is defined to be a closed contour on
the unit circle in the complex $u$-plane,
whereas $\CC_>$
 runs from $0$ to $\infty$ along a ray at an angle $\vphi$
above the negative $x$-axis in the $u$ plane.
The operators on the RHS are defined to have the following
expansions:
\eqn\eperiodic{\eqalign{
\bpm_< (u) &= \pm i \sum_{\om \leq -1/2} \Ga (\ha - \om ) ~ \ma^\om ~
b^\pm_\om u^\om
, ~~~~~\bpm_> (u) = \pm  \sum_{\om \geq 1/2} \frac{2\pi (-1)^{\om + 1/2}}
{\Ga (\om + \ha ) }
 ~ \ma^\om ~ b^\pm_\om u^\om  \cr
\bbpm_< (u) &= \pm  \sum_{\om \leq -1/2} \Ga (\ha - \om ) ~\ma^\om ~ \bbpm_\om
u^{-\om} , ~~~~~
\bbpm_> (u) = \pm i \sum_{\om \geq 1/2} \frac{2\pi (-1)^{\om - 1/2}}
{\Ga (\om + \ha ) }
 ~ \ma^\om ~ \bb^\pm_\om u^{-\om } . \cr}}
One can show
 that the analytic continuation \epercont, \eperiodic\
of \eIIxx\ yields the expansion \eIIIxvii\ of radial quantization.
(See \lec.)

Now, consider the states
\eqn\eIVvi{
\dstate = \inv{(2\pi i)^n} \>
\lvac \bh^{\ep_1} (e^{-i\pi} u_1 ) \cdots \bh^{\ep_n} (e^{-i\pi} u_n )
, ~~~~u_i < 0 . }
Using the analytic continuation \epercont, one can map the above
state into the space of radial quantization.  Bearing in mind that
$u_i <0$ in \eIVvi, one uses the second formula in \epercont\ to
obtain the equation \eIIix, where the vacuum and vertex operators
are:
\eqn\eIVvii{\eqalign{
\lva{\Omega} &= \lvac \cr
V^{\ep} (\th ) &= \inv{2\pi i} \( b^\ep_> (e^{-i\pi} u ) + \bb^\ep_>
(e^{-i\pi} u ) \)  . \cr }}

One can easily check the validity of \eIVvii\ by verifying that it
gives the correct form factors in some simple cases.  As explained
above, any field $\Phi$ in the periodic sector corresponds to a
state $\va{\Phi}$ in the space $\hf^{(p)}$  and
the form-factor is computed as
\eqn\eIVviii{
\dstate \Phi \rangle = \lvac V^{\ep_1} (\th_1 ) \cdots V^{\ep_n} (\th_n )
\va{\Phi } . }
Note that since only the radial annihilation operators appear in
\eIVvii, all of the form factors in the periodic sector have the
`free field' property: for a given field, the $n$-particle form factors
are all zero except for $n=n_{\psi} $, where $n_\psi$ is the number
of free fermion fields needed to construct the field.
Using \eIVviii, \eIVvii, and \eIIIxiv\ one finds
\eqn\eIVix{\eqalign{
{}^{\pm} \lva {\th}  \psi^{\mp} (0) \rvac &= \pm \sqrt{\ma u} ,
{}~~~~~~{}^\pm \lva {\th } \psib^\mp (0) \rvac = \pm i \sqrt{\ma /u} \cr
\tstate J_z (0) \rvac &= -\ma (u_1 u_2 )^{1/2}  , ~~~~~
\tstate J_\zbar (0) \rvac = \ma (u_1 u_2 )^{-1/2}  \cr
\tstate T_{zz} (0) \rvac &= \frac{\ma^2}{2}
\( (u_1)^{1/2} (u_2)^{3/2} - (u_1)^{3/2} (u_2)^{1/2}  \)
\cr
\tstate T_{\zbar\zbar} (0) \rvac &= \frac{\ma^2}{2}
\( (u_1)^{-3/2} (u_2)^{-1/2} - (u_1)^{-1/2} (u_2)^{-3/2}  \)
\cr
\tstate T_{z \zbar} (0) \rvac &= \frac{\ma^2}{2}
\(  \( \frac{u_1}{u_2} \)^{1/2} - \( \frac{u_2}{u_1} \)^{1/2}
\)
. \cr }}
All the higher multiparticle form factors are zero for these
fields, since they are all fermion bilinears.
These form factors agree with expressions derived by the standard
methods in the space of particles $\hp$.

\medskip\noindent
7.2 {\it Anti-Periodic Sector}

In this sector, the analytic continuation of \eIIxx\ that reproduces
the radial expansion \eIIIxvii\ is the following:
\eqn\eIIIxiii{
\int_{-\infty}^\infty \du ~ \bhpm (u)
\rightarrow
\(
\int_{\cL} \dua ~ b^\pm (u) ~~ + ~~
\int_{\cR} \dua ~ \bb^\pm (u) \), }
where
$\cL , \cR$ are contours depending on the angular direction $\vphi$ of
the cut   and
\eqn\eIIIxiv{\eqalign{
\bpm (u) &= \pm i \sum_{\om \in \Zmath } \Gamma (\ha - \om )
\> \ma^\om ~ b^\pm_\om ~ u^\om   \cr
\bbpm (u) &= \pm  \sum_{\om \in \Zmath } \Gamma (\ha - \om )
\> \ma^\om ~ \bb^\pm_\om ~ u^{-\om}  .    \cr }}

The most significant difference between the anti-periodic
and periodic sectors is that here the analytic continuation
\eIIIxiii\ does not separate the radial creation operators from
the annihilation operators, unlike in \epercont.  This simple
fact is what is responsible for the non-free properties of the
form factors in this sector.

Repeating the reasoning given above for the periodic sector, we
propose that again the formula \eIIix\ is valid, where now
\eqn\eIVxii{\eqalign{
\lva{\Omega}  &= \lva{\ha} + \lva{-\ha}  \cr
V^\ep (\th )  &=  \inv{\sqrt{2\pi^2 i } }
\( b^\ep (e^{-i\pi} u )  + \bb^\ep ( e^{-i\pi} u ) \) . \cr}}
 From \eIIIxvi, one sees that the choice \eIVxii\ for
$\lva{\Omega}$ is equivalent to  the following vacuum expectation values:
\eqn\eIVxiv{
\lvac ~ e^{\pm i \phi (0)/2} ~ \rvac = (c\ma )^{1/4} ~
\lva{\Omega} \pm \ha \rangle = (c\ma )^{1/4} . }

One can use the above construction to compute the form factors
of the fields $\exp (\pm i \phi /2 )$.  For these fields,
all of the form factors with a $U(1)$ neutral combination of an
even number of particles is non-zero.  The result is
\eqn\effii{\eqalign{
& {}^{+++...---...} \lva{\th_1 , \th_2 , \cdots , \th_{2n} }
{}~ e^{\pm i \phi (0) /2 } \rvac \cr
&~~~~=
(c\ma )^{1/4}~    \langle \mp \ha \vert
V^+ (u_1 ) \cdots V^+ (u_n ) V^- (u_{n+1} ) \cdots V^- (u_{2n} )
\vert \pm \ha \rangle \cr
&~~~~= (c\ma )^{1/4} \frac{ (\pm 1)^n }{( i\pi )^n } (-1)^{n(n-1)/2}
\sqrt{u_1 \cdots u_{2n} }
\( \prod_{i=1}^n \( \frac{u_{i+n}}{u_i} \)^{\pm 1/2} \)
\( \prod_{i<j \leq n} (u_i - u_j ) \)  \cr
& ~~~~~~~~\times \( \prod_{n+1 \leq i < j } (u_i - u_j ) \)
\( \prod_{r=1}^n \prod_{s=n+1}^{2n} \inv{u_r + u_s } \)   . \cr } }
The above computation can be done using the Wick theorem with the
2-point functions
\eqn\eIVvii{\eqalign{
{}_L \lvacm ~ b^+ (u) \> b^- (u' ) ~ \rvacp_L
= {}_R \lvacp ~ \bb^+ (u) \> \bb^- (u' ) ~ \rvacm_R  &= \pi
\frac{u'}{u+u'} \cr
{}_L \lvacp ~ b^+ (u) \> b^- (u' ) ~ \rvacm_L
= {}_R \lvacm ~ \bb^+ (u) \> \bb^- (u' ) ~ \rvacp_R  &= - \pi
\frac{u}{u+u'} . \cr}}
However the computation is more easily
done using  bosonization techniques
which reasons of space force us to ommit.
After some algebraic manipulation,
one can see that these expressions agree with the known results,
though they were
originally computed using rather different
methods \ref\rmss{\MSS}\rform\foot{The overall numerical factors in
\effii\ differ from the ones in \rform\ however they agree with
results implicit in \rmss. Here the correct normalization is fixed by
the so-called  residue property. }.

\newsec{Conclusions}

Though we have limited ourselves to perhaps the simplest possible
case of the free-fermion point of the sine-Gordon theory, we
believe the ideas presented here can lead to a new framework for
computing form factors in massive integrable quantum field theory.
In this approach, since a complete description of the space of
fields $\hf$ is provided from the outset via radial quantization,
the complete set of solutions to the form factor bootstrap is
automatically yielded.
It is important  however to understand if this is possible more generally.

\vfill\eject

\listrefs
\end